\documentclass[12pt]{article}

\usepackage{amsfonts,latexsym,eucal,color,graphicx,epsfig,amssymb,amsmath}

\newcommand{\be}{\begin{eqnarray}}
\newcommand{\ee}{\end{eqnarray}}
\newcommand{\beq}{\begin{eqnarray}}
\newcommand{\eeq}{\end{eqnarray}}
\newcommand{\pd}{\partial}
\newcommand{\nb}{\nabla}

\newcommand{\dalm}{\kern1pt\vbox{\hrule height 0.9pt\hbox{\vrule width 0.9pt\hskip 2.5pt\vbox{\vskip 5.5pt}\hskip 3pt\vrule width 0.3pt}\hrule height 0.3pt}\kern1pt}
\newcommand{\ie}{{\it i.e.}}
\newcommand{\eg}{{\it e.g.}}
\newcommand{\etc}{{\it etc.}}

\newcommand{\vs}{{\it vs.\ }}

\makeatletter

\@addtoreset{equation}{section}
\makeatletter

\begin{document}

\thispagestyle{empty}

\begin{flushright}
{\small {\tt arXiv:1105.5233 [physics.flu-dyn]}}\\
{\small May 26, 2011}\\
\end{flushright}
\vspace{2cm}
\begin{center}
{\Large {\bf Instability of Compressible Drops and Jets}} \\
\vspace{2cm}
{\bf Umpei Miyamoto}\\
\vspace{.5cm} 
Department of Physics, Rikkyo University, Tokyo 171-8501, Japan\\
\vspace{.5cm}
umpei@rikkyo.ac.jp\\
\vspace{1.5cm}
\end{center}

\begin{abstract}
We revisit the classic problem of the stability of drops and jets held by surface tension, while regarding the compressibility of bulk fluids and spatial dimensions as free parameters. By mode analysis, it is shown that there exists a critical compressibility above which the drops (and disks) become unstable for a spherical perturbation. For a given value of compressibility (and those of the surface tension and density at the equilibrium), this instability criterion provides a minimal radius below which the drop cannot be a stable equilibrium. According to the existence of the above unstable mode of drop, which corresponds to a homogeneous perturbation of cylindrical jet, the dispersion relation of Rayleigh-Plateau instability for cylinders drastically changes. In particular, we identify another critical compressibility above which the homogeneous unstable mode is predominant. The analysis is done for non-relativistic and relativistic perfect fluids, of which self-gravity is ignored.
\end{abstract}


\newpage
\tableofcontents


\section{Introduction}
\label{sec:into}

The free oscillations of liquid droplets were studied by Kelvin~\cite{Kelvin} and Rayleigh~\cite{RayleighSound} more than hundred years ago. Later, Lamb~\cite{Lamb}, Chandrasekhar~\cite{Chandrasekhar}, Reid~\cite{Reid} and others generalized the analysis to take into account the effects of an outer fluid, viscosity, and so on (see, \eg, Introduction of reference \cite{Becker} for a brief but nice review and a more complete list of references). In the simplest case of an inviscid droplet in vacuum (or approximately in air), the droplet with unperturbed radius $r$ and constant density $\rho$ oscillates with the angular frequency given by
\be
	\Omega^2
	=
	\frac{  \sigma }{ \rho r^3 } (\ell-1)\ell(\ell+2) ,
\label{drop-osc}
\ee
where $\ell = 0,1,2,\ldots$ denotes the mode number and $\sigma$ is the surface tension responsible for the oscillation.

Another important phenomenon associated with the surface tension is the drop formation resulting from an instability of cylindrical jets. Theoretical studies date back to the early investigations by Plateau~\cite{Plateau}, Rayleigh~\cite{Rayleigh}, and later by Chandrasekhar~\cite{ChandrasekharBook}  (see \cite{Eggers:1997zz} for a comprehensive review). For example, an inviscid cylinder with non-perturbed radius $r$, the sinusoidal perturbation with wavenumber $k$ evolves in time with the growth rate given by  
\be
	\omega^2
	=
	\frac{  \sigma }{ \rho r^3 }\frac{  kr \left( 1-(kr)^2 \right)  I_1 (kr) }{ I_0 (kr) },
\label{cylind-ins}
\ee
where $I$ is the modified Bessel function of the first kind. This dispersion relation tells us that any cylinder that is longer than $2 \pi r$ is unstable and the most unstable mode, which roughly determines the size of droplets forming, appears at wavelength $ \lambda \sim 9 r $.

The above two phenomena, the oscillations of droplets and the instability of cylinders, are not only of fundamental importance in the theoretical fluid mechanics but also important from industrial points of view. Therefore, they have been studied theoretically and experimentally in variety of physical situations considerable. However, the effect of {\it non-zero compressibility} or {\it finite sound velocity} of fluids has not been well studied, as far as the present author knows. The reason would be that the compressibility of liquids in many non-extremal situations is expected to be negligible and give rise no perceivable effects.

Notwithstanding the above general expectation, in this paper we re-investigate the two classic problems, while allowing the bulk fluids to have non-zero compressibility or finite sound velocity. Our analysis reproduces the results \eqref{drop-osc} and \eqref{cylind-ins} in the incompressible limit. The results are a little surprising and intriguing, at least, at theoretical level. The stability structure of droplets and cylinders with finite compressibility is rather richer than expected. It will be shown that there exists a critical compressibility above which the droplet (or a disk in a two-dimensional case) becomes unstable for a spherically symmetric perturbation. According to the existence of such an instability of disk, a cylinder, whose cross section is the disk, becomes unstable above the critical compressibility. These instability criteria can be interpreted as follow. For given parameters of the fluid and surface, namely the sound velocity $c_s^2 = dp/d\rho$, surface tension, and the density at the equilibrium, there exists a minimum radius of droplet and cylinder below which they cannot be stable equilibria\footnote{If one is familiar with theories of stellar structure, this result may not be totally surprising. That is, a self-gravitating fluid ball is unstable if the heat capacity ratio (or adiabatic index), denoted by $\gamma$ conventionally, exceeds $4/3$ in Newtonian gravitational theory (see, \eg, \cite{198204}). This critical value is changed to a higher value in general relativity (see, \eg, \cite{198204}). Furthermore, if $\gamma$ is slightly larger than 4/3, then the spherical fluid ball is unstable to the radial perturbation provided the radius is less than a critical radius, which is much larger the Schwarzschild radius~\cite{Chandrasekhar2}. The author thanks an anonymous referee for pointing out these points.}. Such a minimum radius is identified to be
\be
	r_{min}
	=
	\frac{n}{n+1}
	\frac{\sigma}{\rho c_s^2},
\ee
where $n=1$ for the cylinders and $n=2$ for the droplets\footnote{We will consider axially symmetric fluids in arbitrary dimensions, by leaving $n$ as a free parameter. The reason to do so is twofold. By leaving the spatial dimension as a free parameter, one can treat the disks, drops, and cylinders at once. Another reason is related to the so-called {\it fluid/gravity correspondence}~\cite{arXiv:0712.2456} (see also \cite{arXiv:0905.4352} for a review and a complete list of references), that was found in the context of the string/M theories and relates the fluid mechanics in $d$ dimensions to a gravitational theory in $(d+1)$ dimensions. In this context, the spatial dimension often has to be treated as a free parameter and some phenomena such as the Rayleigh-Plateau instability and its subsequent dynamics have been known to depend crucially on the dimension~\cite{arXiv:0811.2305,Caldarelli:2008mv}. It is added, however, that within the analysis in this paper, we could not find any qualitative difference originating from the difference of dimension.}. When the fluid is relativistic (\eg, when the pressure is comparable with the energy density $\epsilon$),
\be
	r_{min}
	=
	\frac{n}{n+1}
	\frac{\sigma}{\epsilon c_s^2}
	\left(
		1-(n+1) c_s^2
	\right),
\ee
where
$c_s^2 = dp/d\epsilon$ in this case.

Fortunately (or to the author's regret), in many non-extremal systems such as the water in air at room temperature $r_{min}$ is extremely small ($r_{min}\sim 10^{-11}\;{\rm m}$), and the instability would not play a crucial role in dynamics\footnote{The author confesses that he cannot say for certain whether or not there are systems where $r_{min}$ is macroscopic.
See appendix~\ref{sec:r_min} for discussions on the values of $r_{min}$.}. From a theoretical or mathematical point of view, however, the existence of the minimal radius is significant to prove that the Euler equation supplemented by Young-Laplace's stress balance relation at the surface, which governs the perfect-fluid systems considered in this paper, is not well defined for arbitrary values of $(\sigma, c_s, \rho)$, while these parameters are usually supposed to take any positive finite values.

The organization of this paper is as follow. In section~\ref{sec:non-rel} we review the Euler equation and Young-Laplace relation for non-relativistic fluids, and then we derive the dispersion relation for the perturbations of spherical and cylindrical equilibria. In section~\ref{sec:disp}, we analyze the dispersion relation to obtain the stability criteria for the droplets (section~\ref{sec:k=0}) and cylinders (section~\ref{sec:k>0}). We generalize the analysis to the relativistic fluids in sections \ref{sec:rel} and \ref{sec:disp-rel}. Section~\ref{conc} is devoted to summary and discussion. The stability of the droplets for non-spherical ($\ell \neq 0$) perturbations are shown in appendix~\ref{sec:non-s-pert}.

\section{Non-relativistic perfect fluid with boundary}
\label{sec:non-rel}

\subsection{Euler equation and Young-Laplace relation}
\label{sec:euler}

We consider compressible inviscid fluids in $d$-dimensional ($d \geq 3$) flat spacetime ${\mathbb R^{1,d-1}}$. Denoting density, pressure, and velocity field by $\rho$, $p$, and $v^I$ ($I,J=1,2,\ldots,d-1$), respectively, the continuity equation and Euler equations~\cite{Landau:1987gn} are
\begin{align}
	\pd_t \rho + \nb_I ( \rho v^I )
	&= 0,
\label{cont}
\\
	\pd_t ( \rho v^I ) + \nb_J ( \rho v^I v^J + p g^{IJ} )
	&= 0,
\label{euler}
\end{align}
where $\nb_I$ is the covariant derivative compatible with a flat metric $g_{IJ}$~\cite{Wald:1984rg}. 

We assume that a lump of fluid is supported by a constant surface tension $\sigma$ ($>0$). Then, the Young-Laplace relation~\cite{Landau:1987gn}, describing the normal-stress balance at the surface, is given by
\be
	p = \sigma \kappa \;\big|_{f=0}.
\label{YL}
\ee
Here, $f$ is a scalar function with which the surface is identified by $f=0$, and $\kappa$ is [$(d-2)$-times] the mean curvature of the surface, given as the divergence of the unit normal vector $n^I$ (or as the trace of extrinsic curvature),
\be
	\kappa = \nb_I n^I, \;\;\;
	n_I =  \frac{ \nb_I f }{ ( \nb^J f \nb_J f )^{1/2} }.
\ee
Furthermore, we assume that the surface is convected with the fluid, which is expressed as
\be
	(\pd_t +v^I \nb_I) f = 0 \; \big|_{f=0}.
\label{kbc}
\ee

\begin{figure}[bt]
	\begin{center}
			\includegraphics[width=8cm]{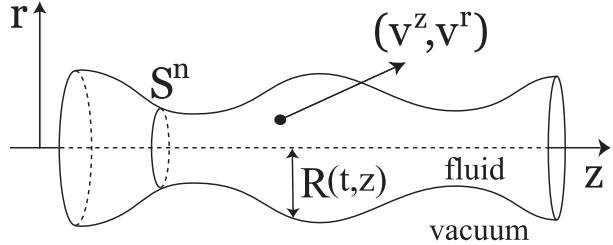}
	\caption{An axially symmetric fluid and the cylindrical coordinates in a ($n+3$)-dimensional spacetime ${\mathbb R^{1,n+2}}$. The fluid surface at $z=Const$ is a $n$-sphere $S^n$.}
	\label{fg:tube}
	\end{center}
\end{figure}

Since we are interested in axially symmetric fluids, it is convenient to work in cylindrical coordinates. Writing the spacetime dimensions as $ d = n+3 $ ($n = 1,2,\ldots$), the line element is written as\footnote{For $d=3$ we just discard the $z$-coordinate, and the cylindrical coordinates reduced to polar coordinates.}
\be
	g_{IJ} dx^I dx^J
	=
	dz^2 + dr^2 + r^2 ds_n^2
	=
	\delta_{ab} dx^a dx^b + r^2 \gamma_{ij}(\theta) d\theta^i d\theta^j,
\label{metric}
\ee
where $x^a := (z,r)$ and $\gamma_{ij}(\theta) d\theta^i d\theta^j$ ($i,j=1,2,\ldots,n$) is the line element on the unit $n$-sphere. Now, we assume that the fluid and its surface have $SO(n+1)$ symmetry around the $z$-coordinate. Then, the coordinate dependence of fluid quantities are given by
\be
	p=p(t,z,r)
\;\;\;
	v^a = v^a (t,z,r),
\;\;\;
	v^i = 0,
\;\;\;
	f(t,z,r) = r - R(t,z),
\ee
where $R(t,z)$ is the function representing the local radius of the fluid surface (see figure~\ref{fg:tube}).

With the above ansatz, continuity equation~\eqref{cont} and Euler equation~\eqref{euler} are written down as
\begin{align}
	( \pd_t + v_z \pd_z + v_r \pd_r ) \rho
	+
	\rho
	\left(
		\pd_z v_z + \pd_r v_r + \frac{n}{r} v_r
	\right) &= 0,
\label{cont_ax}
\\
	\rho ( \pd_t + v_z \pd_z + v_r \pd_r ) v_a
	&=
	- \pd_a p.
\label{euler_ax}
\end{align}
The mean curvature of the surface, appearing in Young-Laplace relation~\eqref{YL}, is given by
\be
	\kappa
	=
	\frac{ n }{ R [ 1+(\pd_z R)^2 ]^{1/2} }
	-
	\frac{ \pd_z^2 R }{ [  1+(\pd_z R)^2 ]^{3/2} }.
\label{kappa_ax}
\ee
Finally, the kinematic boundary condition~\eqref{kbc} is reduced to
\be
	\pd_t R + v_z \pd_z R
	=
	v_r \; \big|_{r=R}.
\label{kbc_ax}
\ee

\subsection{Linear perturbation of cylinders}
\label{sec:pert}

The above system obviously allows the static cylinder as an equilibrium solution, where the constant pressure $p_0$ and the radius of cylinder $R=r_0$ satisfy
\be
	p_0 = \sigma \frac{n}{r_0}.
\label{equi}
\ee

Now, we perturb this equilibrium solution. We can assume that the perturbation results from a sinusoidal disturbance of the local radius given by
\be
	R(t,z)
	=
	r_0
	\left[
		1 + \varepsilon e^{\omega t} \cos (kz)
	\right],
\label{delta_h}
\ee
where $ |\varepsilon|  \ll 1 $ is a small parameter. Such a disturbance leads to those of pressure and velocity, where the disturbed pressure at $O(\varepsilon)$  may take form of
\be
	p(t,z,r)
	=
	p_0 
	\left[
		1 + \varepsilon e^{\omega t} P(r) \cos (kz)
	\right].
\label{delta_p}
\ee
In general, from the perturbations of continuity equation~\eqref{cont} and Euler equation~\eqref{euler}, the pressure perturbation has to satisfy the wave equation
\be
	( \pd_t^2 - c_s^2 \nb^J \nb_J ) \delta p = 0,
\label{wave_eq}
\ee
where $c_s^2 := d p_0 / d \rho_0 $ ($\rho_0$ is the density at the equilibrium and $0<c_s<\infty$) is the sound velocity square of the bulk fluid, and
we denote $O(\varepsilon)$-perturbation of any quantity $X$ by $\delta X$ hereafter. Plugging expression~\eqref{delta_p} into equation~\eqref{wave_eq}, we obtain
\be
	\frac{ d^2 P  }{ d r^2 }
	+
	\frac{n}{r} \frac{ dP }{ dr }
	-
	\left(
		k^2 + \frac{\omega^2}{c_s^2}
	\right) P 
	=
	0.
\ee
With the regularity at the axis ($r=0$), this equation is solved by the modified Bessel function of the first kind
\be
	P (r)
	=
	C \frac
	{
		I_{(n-1)/2}
		\left(  
			 K r
		\right)
	} { r^{(n-1)/2}  },
\;\;\;
	K := \left( k^2 + \frac{\omega^2}{c_s^2} \right)^{1/2},
\label{F}
\ee
where $C$ is an integration constant. The perturbation of Young-Laplace relation~\eqref{YL}, $\delta p = \sigma \delta \kappa |_{r=R} $, fixes the integration constant,
\be
	C
	=
	- \frac{ [ n-(kr_0)^2 ] r_0^{(n-1)/2} }
		   { n I_{(n-1)/2}
			 \left(  
				K r_0
			 \right) 
			}.
\label{C}
\ee

The perturbation of Euler equation in the $r$-direction~\eqref{euler_ax} is $ \rho_0 \pd_t \delta v_r = - \pd_r \delta p $. On the other hand, the velocity in the $r$-direction at the surface is given by $\delta v_r = \pd_t \delta R |_{r=R}$ from kinetic boundary condition \eqref{kbc_ax}. The combination of these two yields
\be
	\pd_r \delta p = - \rho_0 \pd_t^2 \delta R \; |_{r=R}.
\label{delta_ph}
\ee
Plugging equations~\eqref{delta_h}, \eqref{delta_p}, \eqref{F}, and \eqref{C} into \eqref{delta_ph},
and eliminating a derivative of the modified Bessel function, we finally obtain the dispersion relation of perturbations for the compressible cylinder
\be
	\omega^2
	=
	\frac{\sigma}{ \rho_0 r_0^3 }
	[ n-(kr_0)^2 ]
		K r_0
	\frac
	{ I_{(n+1)/2} \left(  K r_0 \right) }
	{ I_{(n-1)/2} \left(  K r_0 \right) }.
\label{disp}
\ee
In the incompressible limit ($c_s \to \infty$), this dispersion relation reduces to equation \eqref{cylind-ins} for $n=1$ and the one derived in \cite{Cardoso:2006sj} for general $n$.

To simplify the dispersion relation, we introduce here the following dimensionless quantities
\be
	\hat{k} := r_0 k,
\;\;\;\;\;
	\hat{\omega} := \left( \frac{\rho_0 r_0^3}{\sigma} \right)^{1/2} \omega,
\;\;\;\;\;
	\beta: = \left( \frac{\sigma}{\rho_0 r_0} \right)^{1/2} c_s^{-1} \;\; (>0).
\label{dimless}
\ee
This $\beta$, being basically the reciprocal of the sound velocity, serves as the parameter representing the (adiabatic) compressibility of the bulk fluid. In terms of these dimensionless quantities, dispersion relation~\eqref{disp} is equivalent to the relation between $\hat{k}$ and $\hat{\omega}$ for which the following function vanishes,
\be
	F(\hat{\omega},\hat{k})
	:=
	\hat{\omega}^2
	-
	(n-\hat{k}^2) \hat{K}
	\frac{ I_{(n+1)/2} ( \hat{K} ) }{ I_{(n-1)/2} ( \hat{K} ) },
\;\;\;
	\hat{K} := \left( \hat{k}^2 + \beta^2 \hat{\omega}^2 \right)^{1/2}.
\label{G}
\ee

\section{Analysis of dispersion relation for the non-relativistic fluid}
\label{sec:disp}

We have derived dispersion relation~\eqref{disp} for the perturbation of cylinders in $ \mathbb{R}^{1,n+2}$. However, since the cross section of a cylinder is a sphere, equation~\eqref{disp} with $k=0$ provides the oscillation frequency (or the growth rate of a possible instability) of droplets in $ \mathbb{R}^{1,n+1}$ (\ie, without the $z$-direction). Thus, we divide this section into two parts, subsections \ref{sec:k=0} and \ref{sec:k>0}. In the former, we investigate the dispersion relation with $k=0$, which corresponds to both a spherically symmetric perturbation of droplets and homogeneous (in the $z$-direction) perturbation of cylinders. Then, in the latter we investigate the dispersion relation for general modes with $k \geq 0$.

\subsection{Instability of drops ($k=0$ mode)}
\label{sec:k=0}

\begin{figure}[t!]
	\begin{center}
		\setlength{\tabcolsep}{ 5 pt }
		\begin{tabular}{ cc }
			(a) & (b) \\
			\includegraphics[width=7.47cm]{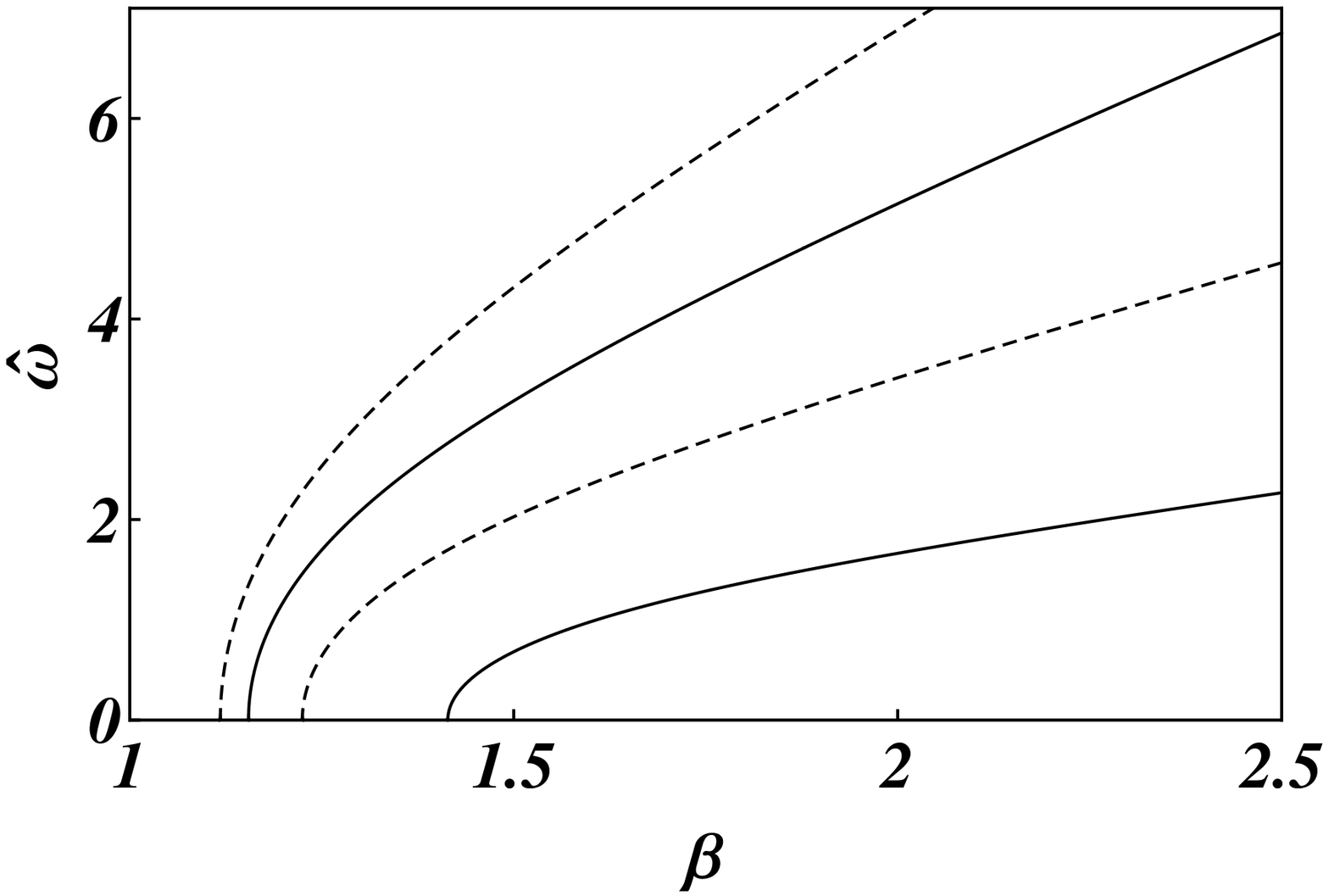} &
			\includegraphics[width=8cm]{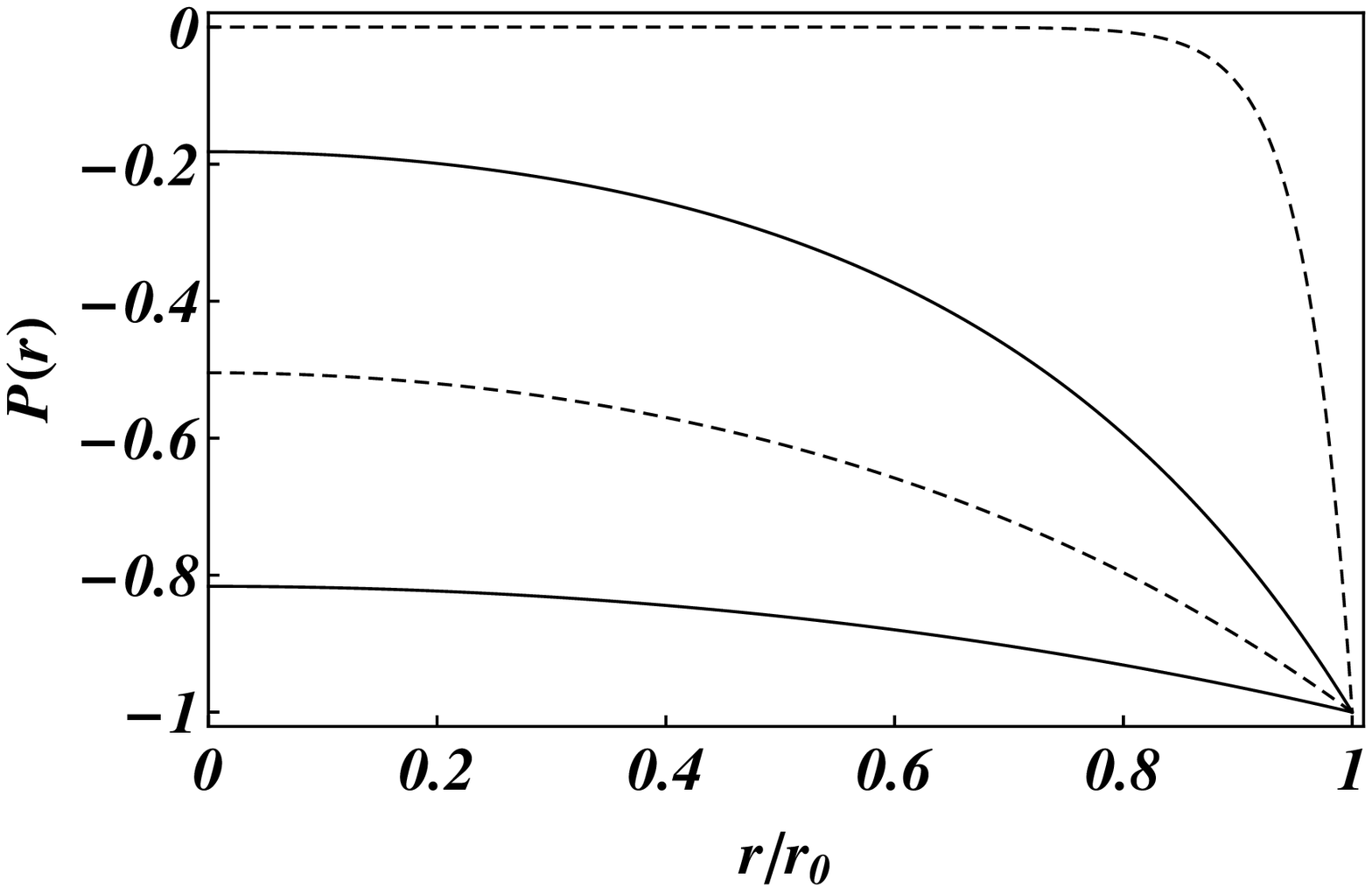} \\
		\end{tabular}
	\caption{(a) Compressibility parameter $\beta$ \vs dimensionless growth rate $\hat{\omega}$ of the radial perturbation of non-relativistic droplets in ${\mathbb R^{1,n+1}}$ for several $n$. $n=1,\; 2, \; 3,\; 4$ from the bottom to the top. (b) The radial eigenfunction $P(r)$ in the $n=2$ case for several values of compressibility parameter. $\beta/\beta_{c,1} = 1.04,\; 1.13,\; 1.30,\; 3.00$ from the bottom curve to the top. Negative $P(r)$ corresponds to the decrease in the pressure, resulting from the increase in the radius of droplet.}
	\label{fg:growth}
	\end{center}
\end{figure}

Setting $k=0$ in \eqref{disp}, we have
\be
	F (\hat{\omega},0)
	=
	\hat{\omega}^2
	-
	n \beta \hat{\omega} \frac{ I_{(n+1)/2} (\beta \hat{\omega}) }
							  { I_{(n-1)/2} (\beta \hat{\omega}) }.
\label{G_0}
\ee
It is noted that in the incompressible limit ($\beta \to 0$), the zero of $F (\hat{\omega},0)$ identically vanishes ($\hat{\omega} = 0$). This just says that the incompressible droplets cannot oscillate nor collapse (nor expand) with keeping the spherical symmetry.

Let us see the behavior of $F(\hat{\omega},0)$ for small $\hat{\omega}$. Expanding equation~\eqref{G_0} around $\hat{\omega} = 0$, one has
\be
	F(\hat{\omega},0)
	=
	\left(
		1-\frac{n}{n+1} \beta^2
	\right) \hat{\omega}^2
	+
	\frac{ n }{ (n+1)^2 (n+3) } \beta^4 \hat{\omega}^4
	+
	O(\hat{\omega}^6).
\label{G_0_exp}
\ee
From this, one can see $F(\hat{\omega},0)|_{\hat{\omega}=0} = \pd_{\hat{\omega}} F(\hat{\omega},0)|_{\hat{\omega}=0} = 0$. On the other hand, it is easy to see that $lim_{\hat{\omega} \to \infty} F(\hat{\omega},0) =  + \infty$ from equation~\eqref{G_0}. Therefore, if $\pd_{\hat{\omega}}^2 F(\hat{\omega},0)|_{\hat{\omega}=0} <0$ holds, $F(\hat{\omega},0)$ must have at least one positive zero from continuity. From equation~\eqref{G_0_exp} it is clear that $\pd_{\hat{\omega}}^2 F(\hat{\omega},0)|_{\hat{\omega}=0} < 0$ holds if $\beta$ is larger than a critical value,
\be
	\beta > \beta_{c,1} := \left( \frac{n+1}{n} \right)^{1/2}.
\label{crit1}
\ee
In this case, from equation~\eqref{G_0_exp} the behavior of $\hat{\omega}$ near $\beta_{c,1}$ is 
\be
	\hat{\omega}
	\simeq
	\left( \frac{4n^3(n+3)^2}{n+1} \right)^{1/4} (\beta-\beta_{c,1})^{1/2}
	+
	O\left( (\beta-\beta_{c,1})^{3/2} \right).
\ee

The global $\beta$-dependence of $\hat{\omega}$ obtained from equation~\eqref{G_0} is shown in figure~\ref{fg:growth}(a).
One can see that in the large-$\beta$ limit, $\hat{\omega}$ increases linearly with $\beta$. This asymptotic behavior, that is in fact $\hat{\omega} \simeq n \beta $, can be derived from equation~\eqref{G_0} with using a property of modified Bessel function, $lim_{\hat{\omega} \to \infty} I_{(n+1)/2}(\beta \hat{\omega})/I_{(n-1)/2}(\beta \hat{\omega}) = 1$. Radial function $P(r)$ in the $n=2$ case for several values of $\beta$ is shown in figure~\ref{fg:growth}(b). One can observe the non-uniform (in the $r$-direction) decrease in the pressure, resulting from the increase in the droplet radius. The non-uniformity of the pressure perturbation is amplified as the compressibility increases, which is interpreted that the compressibility reduces the propagation speed of density fluctuations generated near the surface. 

With using equations~\eqref{equi} and \eqref{dimless}, the instability criterion \eqref{crit1} can be written in several forms. Namely, $ c_s^2 < p_0 /[(n+1)\rho_0] $ or equivalently
\be
	r_0
	<
	r_{min}
	:=
	\frac{n}{n+1} \frac{\sigma}{\rho_0 c_s^2}.
\label{crit3}
\ee
Inequality~\eqref{crit3} is striking, which means that there exists a minimum radius $r_{min}$ only above which the droplet and cylinder can exist stably. The critical radius is determined by three parameters ($\rho_0,\sigma,c_s$), which do not restrict each other at least from a macroscopic point of view, although should be correlated microscopically\footnote{The value of $r_{min}$ for the water in air at temperature $25 \; {}^\circ\mathrm{C}$ is around $ 2.13 \times 10^{-11} \; {\rm m}$ (estimated with $n=2$, $\sigma = 72.0 \times 10^{-3}\; {\rm J/m^2}$, $\rho_0 = 1.00 \times 10^3 \;{\rm kg/m^3} $, and $ c_s = 1.50 \times 10^{3} \; {\rm m/s} $ \cite{Data}), where the fluid-mechanical description has already broken down. Thus, the instability found plays no central role for such a fluid. It would be interesting to look for a system in which $r_{min}$ is larger than or comparable with the length scale where the fluid approximation breaks down. See also appendix \ref{sec:r_min}.}.

One could consider non-spherical perturbations of the drop by allowing $\theta$-dependence of $R$ in equation~\eqref{delta_h}. As in the incompressible case, however, the non-spherical modes turn out to be oscillatory for the compressible fluids too, proving that the droplets are stable for such perturbations. See appendix~\ref{sec:non-s} for a proof.

\subsection{Rayleigh-Plateau instability ($k > 0$ modes)}
\label{sec:k>0}

We proceed the dispersion relation~\eqref{disp} for general values of $k \geq 0$. Since the homogeneous ($k=0$) mode becomes unstable ($ \omega > 0 $) above the critical compressibility $\beta_{c,1}$, one can expect the behavior of the dispersion relation for $k>0$ changes at the critical compressibility.

\begin{figure}[h!]
	\begin{center}\includegraphics[width=8cm]{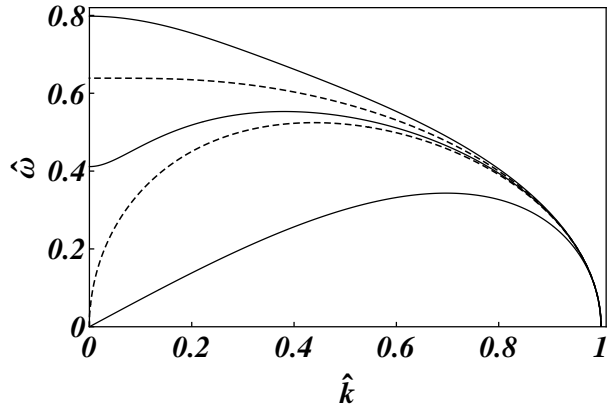}
	\caption{Dimensionless wavenumber $\hat{k}$ \vs dimensionless growth rate $\hat{\omega}$ of the instability of non-relativistic cylinder in ${\mathbb R^{1,3}}$ ($n=1$) for several values of compressibility. $\beta = 0.00, \; 1.41 \; (=\beta_{c,1}), \; 1.45, \; 1.49 \;(=\beta_{c,2}), \; 1.53 $ from the bottom to the top.}
	\label{fg:disp}
	\end{center}
\end{figure}

The dispersion relation~\eqref{disp} for $n=1$ is shown in figure~\ref{fg:disp} for several values of $\beta$. Qualitative behaviors are independent of $n$. For $ 0  < \beta \leq \beta_{c,1}$, the dispersion relation is qualitatively the same as the usual dispersion relation of Rayleigh-Plateau instability (for an incompressible inviscid fluid), although the growth rate for all $ \hat{k} \in ( 0,\sqrt{n} )$ mildly increases with $\beta$. The wavenumber and growth rate of the most unstable mode at $\beta=\beta_{c,1}$ obtained numerically are given in table~\ref{tbl:crit}. As $\beta$ exceeds $\beta_{c,1}$, the homogeneous ($k=0$) mode begins to have positive growth rate as shown in section~\ref{sec:k=0}. Incidentally, the dispersion relation deviates from that of usual Rayleigh-Plateau instability. Note that even if $\beta$ exceeds $\beta_{c,1}$ only slightly, the most unstable mode is still in $k>0$. However, there exists another critical value of $\beta$ [we call it $\beta_{c,2}$ ($ >\beta_{c,1} $)] above which the homogeneous mode becomes the most unstable one. The values of $\beta_{c,2}$ and growth rate of the most unstable mode therein, that we denote by $\hat{\omega}_{\ast,2}$, are also given in table~\ref{tbl:crit}. They can be obtained by solving numerically the following coupled algebraic equations for $\beta$ and $\hat{\omega}$,
\be
	F(\hat{\omega},\hat{k})|_{\hat{k}=0} = 0,
\;\;\;
	\pd_{\hat{k}}^2 F(\hat{\omega},\hat{k})|_{\hat{k}=0} = 0.
\ee 
We note that $\beta_{c,2}$ is of course larger than $\beta_{c,1}$ but only slightly. Thus, we can say that for generic values of $\beta$ larger than $\beta_{c,1}$ the homogeneous unstable mode is predominant.

\begin{table}[h!]
\begin{center}
\caption{The critical values of compressibility parameter, $\beta_{c,1}$ and $\beta_{c,2}$, and characteristic dimensionless wavenumber and growth rate. $\beta_{c,1}$ is the compressibility above which the cylinder is unstable for the homogeneous ($k=0$) perturbation. $\hat{k}_{\ast,1}$ and $\hat{\omega}_{\ast,1}$ are the wavenumber and growth rate of the most unstable mode when $\beta=\beta_{c,1}$, respectively. $\beta_{c,2}$ is the compressibility above which the homogeneous ($k=0$) mode becomes the most unstable one. $\hat{\omega}_{\ast,2}$ is the largest growth rate at $\beta=\beta_{c,2}$.}
\label{tbl:crit}
\vspace{2mm}
\setlength{\tabcolsep}{8 pt}
\begin{tabular}{ccccccccccccccccc}
\hline\hline
$n$ & 1 & 2 & 3  & 4 & 5 & 6 & 7 & 8 & 9 & 10 \\
\hline
$\beta_{c,1} $  & 1.414 & 1.225 & 1.155 & 1.118 & 1.095 & 1.080 & 1.069 & 1.061 & 1.054 & 1.049 \\
$\hat{k}_{\ast,1}$  & .4388 & .6252 & .7596 & .8672 & .9580 & 1.037 & 1.107 & 1.171 & 1.229 & 1.282 \\
$\hat{\omega}_{\ast,1}$  & .5244 & .8504 & 1.115 & 1.345 & 1.552 & 1.742 & 1.919 & 2.086 & 2.244 & 2.394 \\
$\beta_{c,2} $  & 1.489 & 1.293 & 1.215 & 1.172 & 1.144 & 1.124 & 1.110 & 1.098 & 1.089 & 1.081 \\
$\hat{\omega}_{\ast,2}$  & .6388 & 1.039 & 1.358 & 1.632 & 1.875 & 2.097 & 2.301 & 2.492 & 2.671 & 2.840 \\
\hline\hline
\end{tabular}
\end{center}
\end{table}

\section{Relativistic perfect fluid with boundary}
\label{sec:rel}

The argument in the preceding sections can be generalized to relativistic fluids. We consider relativistic fluids in the $d$-dimensional flat spacetime $\mathbb{R}^{1,d-1}$ ($d \geq 3$) with the spacetime coordinates $x^\mu = (ct, x^I)$ ($\mu,\nu = 0,1,2,\ldots,d-1$; $I,J=1,2,\ldots,d-1$), and denoting the flat metric by $g_{\mu\nu}$ [with the so-called almost-plus notation ($-,+,\cdots,+$)]. The speed of light is set to unity ($c=1$). Symbols appearing hereafter have the same meanings as in the non-relativistic case, otherwise noted.

\subsection{Euler equation and Young-Laplace relation}
\label{sec:euler-rel}

The energy-momentum tensor of a relativistic perfect fluid held by surface tension (see, \eg, \cite{Misner:1974qy,Lahiri:2007ae}) is given by
\be
	T^{\mu\nu}
	=
	( \epsilon u^\mu u^\nu + p P^{\mu\nu} ) \Theta (-f) 
	-
	\sigma | \nb f | h^{\mu\nu} \delta (f).
\ee
Here, $\epsilon$ is the energy density; $u^\mu$ the normalized $d$-velocity field ($u^\mu u_\mu = -1)$; $\Theta$ the Heaviside step function; $\delta$ the Delta function;
$ P^{\mu\nu} := g^{\mu\nu} + u^\mu u^\nu $ and $ h^{\mu\nu} := g^{\mu\nu} - n^\mu n^\nu $ are the projection tensors; $n^\mu$ is the unit normal of the surface defined by $ n_\mu = \nb_\mu f / (\nb^\nu f \nb_\nu f)^{1/2} $; $u^\mu$ and $n^\mu$ are orthogonal each other at the surface $u^\mu n_\mu = 0 |_{f=0}$.

Projecting the energy-momentum conservation $\nb_\mu T^{\mu\nu} = 0$ onto $u^\mu$ and $P^{\mu\nu}$, we obtain the relativistic continuity and Euler equations,
\begin{align}
	u^\alpha \nb_\alpha \epsilon + (\epsilon+p) \nb_\alpha u^\alpha
&
	=
	0,
\label{cont_rel}
\\
	( \epsilon+p ) u^\alpha \nb_\alpha u^\mu
&
	=
	- P^{\mu \alpha} \nb_\alpha p.
\label{euler_rel}
\end{align}
The Young-Laplace relation, obtained by projecting the surface contribution of the energy-momentum conservation onto $n^\mu$, takes the same form as the non-relativistic case~\eqref{YL}, but the mean curvature in the this case is given by the $d$-dimensional divergence,
\be
	\kappa =  \nb_\mu n^\mu.
\label{kappa_rel}
\ee
The kinematic boundary condition is given by
\begin{align}
	u^\alpha \nb_\alpha f
	=
	0 \; \big|_{f=0}.
\label{kbc_rel}
\end{align}

As in the non-relativistic case, we introduce cylindrical coordinates in which the line element of flat spacetime is given by\footnote{For $d=3$ we just discard the $z$-coordinate as in the non-relativistic argument.}
\be
	g_{\mu\nu} d x^\mu dx^\nu
	=
	- dt^2 + dz^2 + dr^2 + r^2 ds_n^2
	=
	\eta_{ab} dx^a dx^b + r^2 \gamma_{ij}(\theta) d\theta^id\theta^j,
\label{metric}
\ee
where $x^a = (t,z,r)$ and $\eta_{ab} = Diag.~(-1,1,1)$ is the three-dimensional flat Lorentzian metric. Assuming that the fluid and surface are axially symmetric around the $z$-axis, the coordinate dependence of the pressure, velocity field, and surface are
\be
	p = p(t,z,r),
\;\;\;
	u^a = u^a (t,z,r),
\;\;\;
	u^i = 0,
\;\;\;
	f = r-R(t,z).
\ee

Continuity equation \eqref{cont_rel} and Euler equation~\eqref{euler_rel} are written down as\footnote{
In the present coordinates the non-vanishing components of Christoffel symbol are
$ \Gamma^{a}_{ij} = - r \delta^a_r \gamma_{ij}$, $\Gamma^i_{ja} = r^{-1} \delta^i_j \delta^r_a$, and $ \Gamma^i_{jk} = {}^{(\gamma)}\Gamma^i_{jk},$ where ${}^{(\gamma)}\Gamma^i_{jk}$ is the Christoffel symbol with respect to $\gamma_{ij}$. With using these, the vector derivatives are calculated. For example, $ \nb_a u_b = \pd_a u_b$, $ \nb_i u_j = r u^r \gamma_{ij}$, and $ \nb_\alpha u^\alpha = \pd_a u^a + nr^{-1} u^r$.}
\begin{align}
	u^a \pd_a \epsilon
	+  
	( \epsilon + p )
	\left(
		\pd_a u^a + \frac{n}{r} u^r
	\right)
	&=
	0,
\label{cont_rel_ax}
\\
	(\epsilon+p) u^b \pd_b u_a
	&=
	-P_a^b \pd_b p,
\label{euler_rel_ax}
\end{align}
where the indices ($a,b,\ldots$) are raised and lowered by $\eta^{ab}$ and $\eta_{ab}$, respectively. The mean curvature \eqref{kappa_rel}, appearing in the Young-Laplace relation~\eqref{YL}, is given by
\begin{multline}
	\kappa
	=
	\frac{ n }{ R [ 1-(\pd_t R)^2 + (\pd_z R)^2 ]^{1/2} }
	-
	\frac{ [1-(\pd_t R)^2] \pd_z^2 R - [1+(\pd_z R)^2] \pd_t^2 R + 2(\pd_t R)(\pd_t \pd_z R)\pd_z R }
		 { [ 1-(\pd_t R)^2 + (\pd_z R)^2 ]^{3/2} }.
\label{kappa_full}
\end{multline}
Finally, kinetic boundary condition \eqref{kbc_rel} reads
\be
	u^t \pd_t R + u^z \pd_z R = u^r \; \big|_{r=R}.
\label{kbc_rel_ax}
\ee

\subsection{Linear perturbation of cylinders}
\label{sec:pert-rel-ax}

In general, perturbing relativistic continuity equation \eqref{cont_rel} and Euler equation \eqref{euler_rel} around a static equilibrium where $(p,\epsilon,u^\mu) = (p_0,\epsilon_0,\delta^\mu_t)$, we have
\begin{align}
	\pd_t \delta p  + c_s^2 ( \epsilon_0+p_0 ) \nb_\alpha \delta u^\alpha 
&
	= 0,
\label{cont_pert_rel}
\\
	( \epsilon_0 + p_0 ) \pd_t \delta u^\mu  +  P^{\mu\alpha}_{(0)} \nb_\alpha \delta p
\label{euler_pert_rel}
&
	= 0,
\end{align}
where $P^{\mu\nu}_{(0)} := g^{\mu\nu} + \delta^\mu_t \delta^\nu_t$ and $c_s^2 := d p_0 / d \epsilon_0 $ is the sound velocity square (note that $ c_s < 1$ from the causality).
Eliminating $\delta u^\mu$ from these two equations, one obtains an wave equation for the pressure perturbation,
\be
	\left( \pd_t^2 - c_s^2 P^{\alpha\beta}_{(0)} \nb_\alpha \nb_\beta \right) \delta p = 0.
\label{wave_eq_rel}
\ee

As in the non-relativistic case in section~\ref{sec:pert}, the axially symmetric relativistic system in section~\ref{sec:euler-rel} allows the cylinder as a static equilibrium, where constant pressure $p_0$ and radius $R=r_0$ satisfy equation~\eqref{equi}. Plugging ansatz \eqref{delta_p} into wave equation \eqref{wave_eq_rel} and using the perturbation of Young-Laplace relation $\delta p = \sigma \delta \kappa |_{r=R}$, we obtain the radial function in the relativistic case,
\be
	P(r)
	=
	- 
	\frac{ [ n-( k^2 + \omega^2 ) r_0^2 ] r_0^{(n-1)/2} }
		 { n I_{(n-1)/2} ( Kr_0 ) } 
	\frac{ I_{(n-1)/2} (Kr) }{ r^{(n-1)/2} },
\;\;\;
	K := \left( k^2 + \frac{ \omega^2 }{ c_s^2 } \right)^{1/2}.
\label{F_rel}
\ee

The perturbation of Euler equation \eqref{euler_pert_rel} in the $r$-direction reads $ (\epsilon_0+p_0) \pd_t \delta u^r + \pd_r \delta p = 0 $. On the other hand, the perturbation of kinetic boundary condition \eqref{kbc_rel_ax} reads $ \pd_t \delta R = \delta u^r |_{r=R} $. Eliminating $\delta u^r$ from these two equations, we obtain
\be
	( \epsilon_0 + p_0 ) \pd_t^2 \delta R
	=
	- \pd_r \delta p \;\big|_{r=R}.
\label{delta_hp_rel}
\ee
Plugging equations~\eqref{delta_h}, \eqref{delta_p}, and \eqref{F_rel}, into \eqref{delta_hp_rel}, we obtain the dispersion relation for the perturbation of relativistic compressible cylinder
\be
	\omega^2 = 
	\frac{ \sigma }{ ( \epsilon_0 + p_0 ) r_0^3 }
	[ n-( k^2 + \omega^2 )r_0^2 ] Kr_0 
	\frac{ I_{(n+1)/2} ( Kr_0 ) }{ I_{(n-1)/2} ( Kr_0 ) }.
\label{disp_rel}
\ee
It is noted that a similar result for a relativistic compressible fluid with a particular equation of state was obtained in~\cite{Caldarelli:2008mv}. 

We introduce the following dimensionless quantities (remember that we have already set the speed of light to unity),
\be
	\hat{k} := r_0 k,
\;\;\;\;\;
	\hat{\omega} := r_0 \omega,
\;\;\;\;\;
	\hat{\sigma} := (\epsilon_0 r_0)^{-1} \sigma,
\;\;\;\;\;
	\beta := c_s^{-1} \;(>1).
\label{dimless_rel}
\ee
Then, dispersion relation \eqref{disp_rel} is the relation between $\hat{k}$ and $\hat{\omega}$ for which the following function vanishes,
\be
	F(\hat{\omega},\hat{k})
	:=
	\hat{\omega}^2
	-
	\frac{\hat{\sigma}}{1+n\hat{\sigma}}
	[ n- (\hat{k}^2+\hat{\omega}^2) ] \hat{K}
	\frac{ I_{(n+1)/2}(\hat{K})  }
		 { I_{(n-1)/2}(\hat{K}) },
\;\;\;
	\hat{K} := \left( \hat{k}^2 + \beta^2 \hat{\omega}^2 \right)^{1/2}.
\label{G_rel}
\ee
Function $F(\hat{\omega},\hat{k})$ depends on not only the dimensionless compressibility $\beta$ but also the dimensionless surface tension $\hat{\sigma}$. This is contrast to the non-relativistic counterpart \eqref{G}, that depends only on the compressibility $\beta$.

\section{Analysis of dispersion relation for the relativistic fluid}
\label{sec:disp-rel}

As explained at the beginning of section~\ref{sec:k=0}, the homogeneous ($k=0$) mode has the special meaning that it corresponds to both the homogeneous perturbation of the cylinders in ${\mathbb R^{1,n+2}}$ and the spherical perturbation of droplets in ${\mathbb R^{1,n+1}}$. In the first subsection below we look into the $k=0$ mode, then we proceed the analysis of general $k \geq 0$ perturbations in the second subsection.

\subsection{Instability of drops ($k=0$ mode)}
\label{sec:k=0-rel}

\begin{figure}[t!]
	\begin{center}
		\setlength{\tabcolsep}{ 10 pt }
		\begin{tabular}{ cc }
			(a) & (b) \\
			\includegraphics[width=7.8cm]{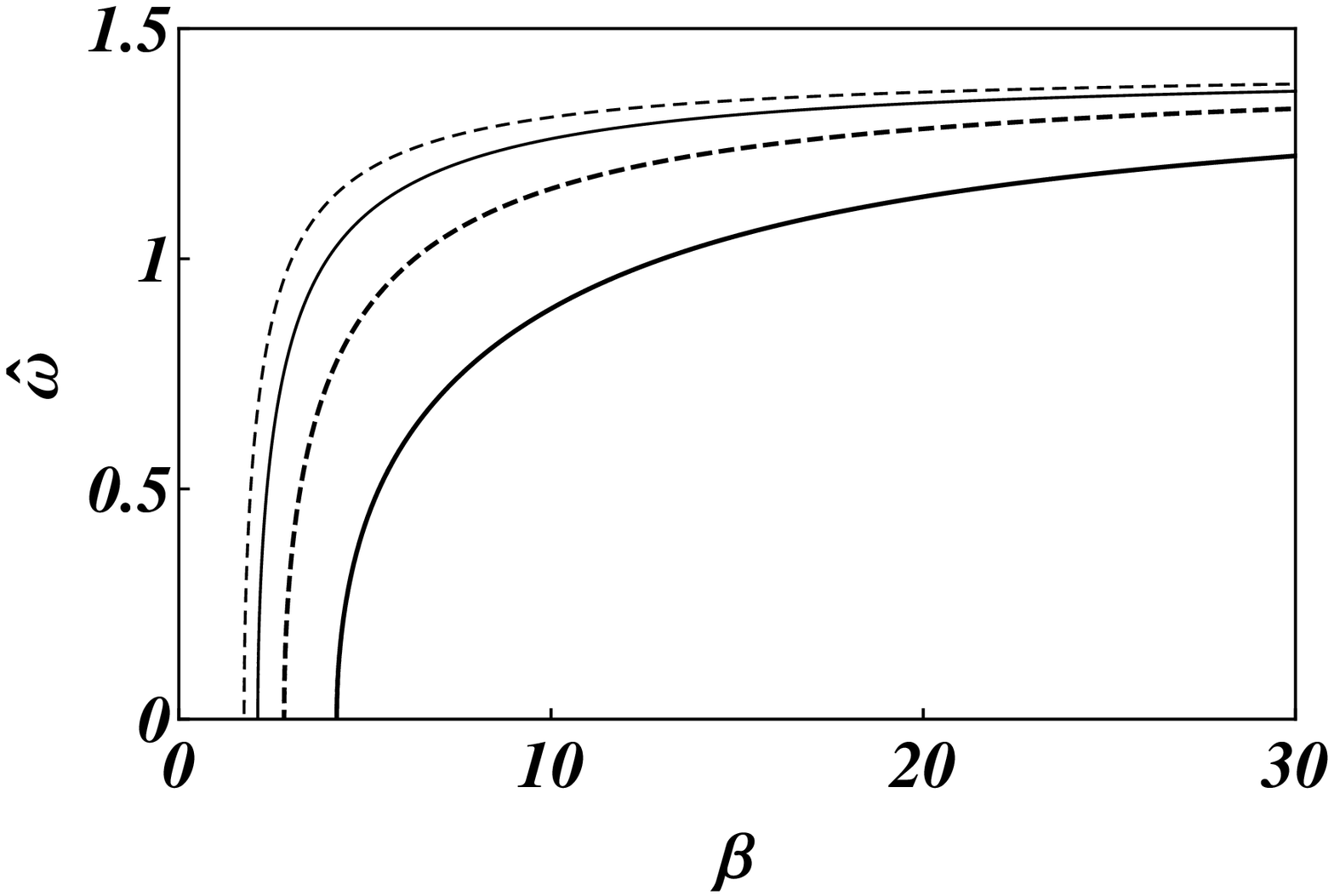}
			&
			\includegraphics[width=8cm]{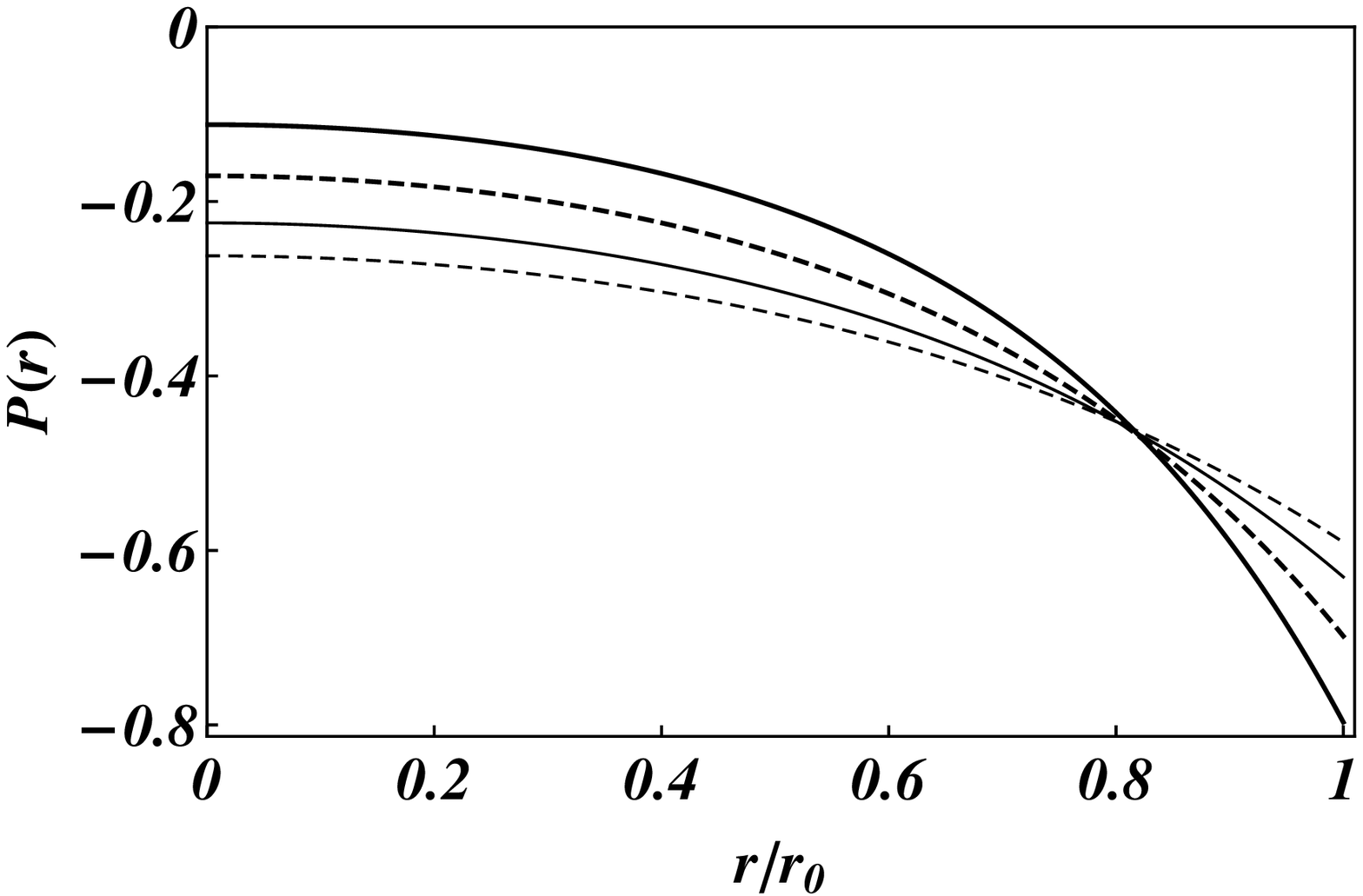} \\
		\end{tabular}
	\caption{(a) Compressibility parameter $\beta$ \vs dimensionless growth rate of the radial perturbation of droplet in ${\mathbb R^{1,3}}$ ($n=2$) for several values of dimensionless tension. $\hat{\sigma} = 0.100$ (thick solid) $0.300$ (thick dashed), $1.00$ (thin solid), $20.0$ (thin dashed). (b) Radial function $P(r)$ for the same set of $\hat{\sigma}$ as that in figure~\ref{fg:growth_rel}(a). $n=2$ and $\beta/\beta_{c,1} = 1.50$.}
	\label{fg:growth_rel}
	\end{center}
\end{figure}

Setting $k=0$ in equation~\eqref{G_rel}, we have
\be
	F(\hat{\omega},0)
	=
	\hat{\omega}^2
	+
	\frac{\hat{\sigma}}{1+n\hat{\sigma}}
	(\hat{\omega}^2-n) \beta \hat{\omega}
	\frac{ I_{(n+1)/2} (\beta \hat{\omega}) }{ I_{(n-1)/2} (\beta \hat{\omega}) }.
\label{G0_rel}
\ee
Let us see the behavior of $F(\hat{\omega},0)$ for small $\hat{\omega}$ by expanding it around $\hat{\omega}=0$, 
\be
	F(\hat{\omega},0)
	=
	\left(
		1 - \frac{ n\hat{\sigma} }{ (n+1)(1+n\hat{\sigma}) } \beta^2
	\right) \hat{\omega}^2
	+
	\frac{  n \beta^2 + (n+1)(n+3)  }
		 { (n+1)^2 (n+3) (1+n\hat{\sigma}) } \hat{\sigma} \beta^2 \hat{\omega}^4
	+
	O(\hat{\omega}^6).
\label{G0_exp_rel}
\ee
From this, one can see $F(\hat{\omega},0)|_{\hat{\omega}=0} = \pd_{\hat{\omega}} F(\hat{\omega},0)|_{\hat{\omega}=0} = 0$. On the other hand, one can see $lim_{\hat{\omega} \to \infty} F(\hat{\omega},0) = + \infty$ from equation~\eqref{G0_rel}. Thus, if $\pd_{\hat{\omega}}^2 F(\hat{\omega},0)|_{\hat{\omega}=0} < 0$ holds, $ F(\hat{\omega},0) $ must have at least one positive zero from continuity. From equation~\eqref{G0_exp_rel}, one can see that $\pd_{\hat{\omega}}^2 F(\hat{\omega},0)|_{\hat{\omega}=0} < 0$ holds if the compressibility parameter is greater than a critical value,
\be
	\beta
	>
	\beta_{c,1}
	:=
	\left(
		\frac{ (n+1)(1+n\hat{\sigma}) }{ n\hat{\sigma} }
	\right)^{1/2}.
\label{rel_crit1}
\ee
In this case, the $\beta$-dependence of growth rate near $\beta = \beta_{c,1}$ can be read from equation~\eqref{G0_exp_rel},
\be
	\hat{\omega}
	\simeq
	\left(
		\frac{ 4n^3 (n+3)^2  \hat{\sigma}^3}
			 { (n+1)(1+n\hat{\sigma})[ 1+(2n+3)\hat{\sigma} ]^2 }
	\right)^{1/2}
	( \beta - \beta_{c,1} )^{1/2}
	+ O
	\left(
		(\beta-\beta_{c,1} )^{3/2}
	\right).
\ee
The global behavior of $\hat{\omega}(\beta)$ for several values of $\hat{\sigma}$ is shown in figure~\ref{fg:growth_rel}(a). In the present relativistic case, in contrast to the non-relativistic case, the growth rate asymptotes to a constant in the large-$\beta$ limit, that is $\hat{\omega} \simeq \sqrt{n}$ in fact. This can be derived from equation~\eqref{G_rel}. Radial function $P(r)$ is shown in figure~\ref{fg:growth_rel}(b).  

Getting back to dimensionful quantities with equation~\eqref{dimless_rel}, instability criterion \eqref{rel_crit1} is rewritten as
\be
	r_0
	<
	r_{min}
	:=
	\frac{n}{n+1}  \frac{\sigma}{ \epsilon_0 c_s^2 }
	\left(
		1-(n+1) c_s^2
	\right).
\label{rel_crit2}
\ee
Namely, there exists minimum radius $r_{min}$ below which the drops and cylinders become unstable. See appendix \ref{sec:non-s_rel} for the proof of stability for non-spherical perturbations. 

\begin{figure}[h!]
	\begin{center}
		\setlength{\tabcolsep}{ 10 pt }
		\begin{tabular}{ cc }
			(a) & (b) \\
			\includegraphics[width=7.72cm]{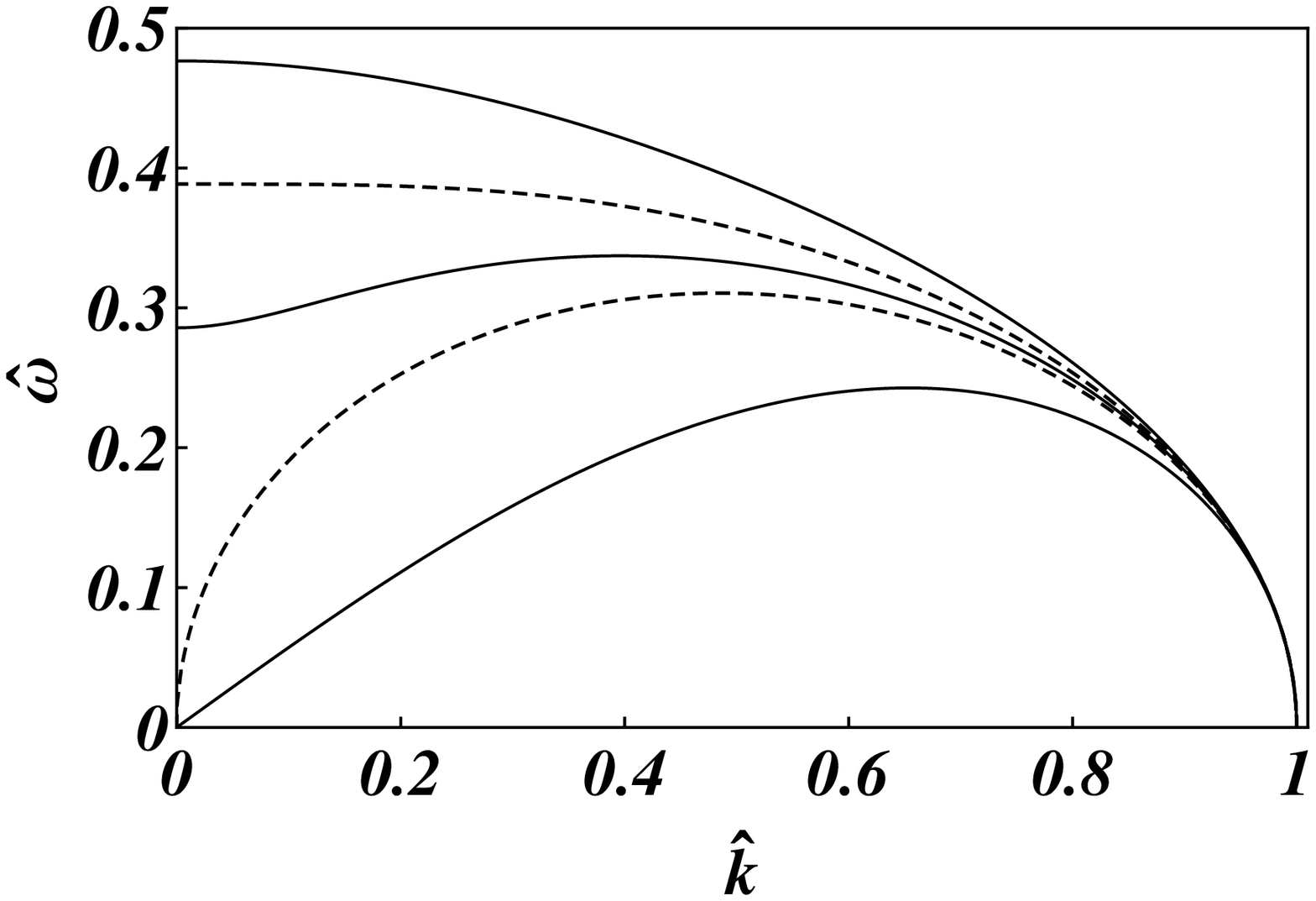}
			&
			\includegraphics[width=8cm]{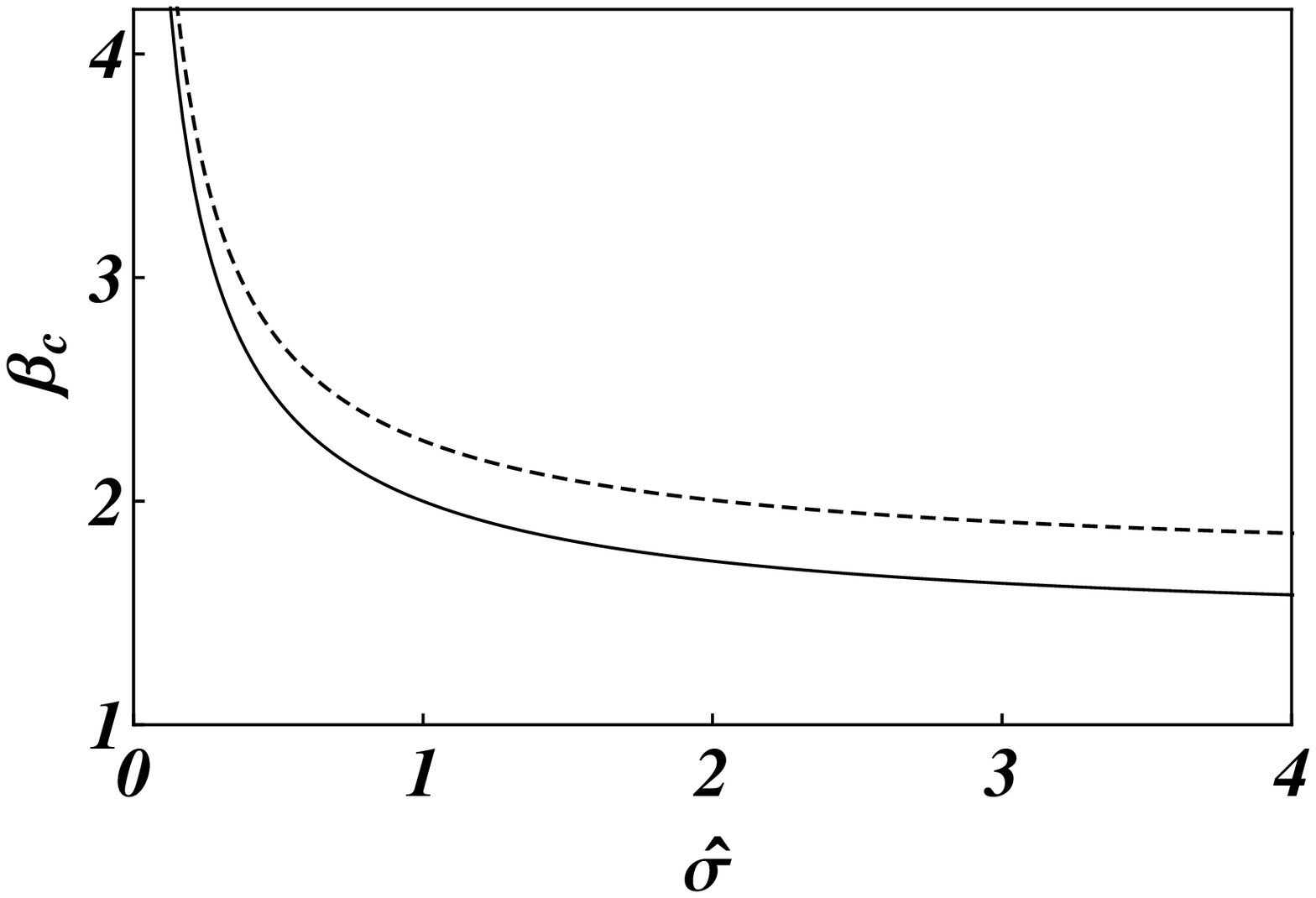} \\
		\end{tabular}
	\caption{(a) Dimensionless wavenumber $\hat{k}$ \vs dimensionless growth rate $\hat{\omega}$ of the perturbation of relativistic cylinder in ${\mathbb R^{1,3}}$ ($n=1$) for several values of compressibility.  $\beta = 1.00, \; 2.00 \; (=\beta_{c,1}), \; 2.13, \; 2.27 \;(=\beta_{c,2}), \; 2.45 $ from the bottom to the top. The dimensionless surface tension is fixed to $\hat{\sigma}=1$. (b) The $\hat{\sigma}$-dependence of the critical compressibility, $\beta_{c,1}$ (solid) and $\beta_{c,2}$ (dashed), for $n=1$.}
	\label{fg:dispRel}
	\end{center}
\end{figure}

\subsection{Rayleigh-Plateau instability ($k > 0$ modes)}
\label{sec:k>0-rel}

The behavior of dispersion relation $\hat{\omega}(\hat{k})$ is quite similar to the non-relativistic case, except that the critical values of compressibility, $\beta_{c,1}$ and $\beta_{c,2}$, depend on $\hat{\sigma}$. Numerical plot of $\hat{\omega}=\hat{\omega}(\hat{k})$ for several values of $\beta$ is shown in figure~\ref{fg:dispRel}(a). The value of $\hat{\sigma}$ does not affect the qualitative behavior of the dispersion relation. In order to see that the second critical compressibility $\beta_{c,2}$, above which the $k=0$ mode is the most unstable one, is only slightly larger than the first critical value $\beta_{c,1}$ in all range of $\hat{\sigma}$, we numerically plot $\beta_{c,1}$ and $\beta_{c,2}$ for $n=1$ in figure~\ref{fg:dispRel}(b). 

\section{Conclusion}
\label{conc}

We have investigated the stability of spherical drops and cylindrical jets held by the surface tension, in particular, its dependence on the compressibility $\beta$ or sound velocity $c_s$ of the bulk fluids. For simplicity, we have focused on the perfect fluids (\ie, inviscid fluids with no heat transfer) immersed in vacuum, while we consider both the non-relativistic and relativistic fluids in general dimensions, which allow us to treat the disk, droplets, and cylindrical jets simultaneously in a systematic way.

As the main result, we have shown that there exists the critical compressibility $\beta_{c,1}$ for both the non-relativistic and relativistic fluids, equations \eqref{crit1} and \eqref{rel_crit1}, above which the spherical drops are unstable for the spherical perturbation. For given parameters of the fluid and surface, \ie, the surface tension $\sigma$, and the sound velocity $c_s$ and density at the equilibrium $\rho_0$ (or $\epsilon_0$ in the relativistic case), the instability criteria poses the lower limit on the droplet size $r_{min} \sim \sigma/(\rho_0 c_s^2)$ [see equations \eqref{crit3} and \eqref{rel_crit2}], below which any droplets cannot be stable equilibria.

We have shown also that according to the instability of disks and droplets, which corresponds to the instability of cylinders for homogeneous perturbations, the dispersion relation of Rayleigh-Plateau instability exhibits the significant change. Namely, for $ \beta > \beta_{c,1} $ the cylinders are unstable for the perturbations that are homogeneous in the axial direction. Furthermore, such a mode becomes the most unstable one above the second critical compressibility $\beta_{c,2}$, which is slightly larger than $\beta_{c,1}$ in general.

Here, let us stress the significance of minimum radius $r_{min}$. In the framework of fluid mechanics, it has been assumed or simply believed that {\it any} positive finite values can be given to the three quantities $\sigma$, $c_s$, and $\rho_0$ (although they should be correlated each other if one pursues their origins from a microscopic point of view). We have shown, however, that the spherical droplets, which are the most fundamental equilibrium states of localized fluids in fact, exhibit the instability for $r_0 < r_{min} \sim \sigma/(\rho_0 c_s^2)$. Therefore, one cannot give values to the three parameters freely in order to describe arbitrarily small-scale dynamics successfully. In other words, the systems defined by the Euler equation and Young-Laplace relation intrinsically contain the instability, and are not well defined in certain regimes of parameter space.

We have adopted several assumptions for simplicity, such as the absence of viscosities, heat transfer, and outer fluids, the constancy of surface tension, and so on. In addition, the instability discovered is just the result of mode analysis, that can never predict the following dynamics. Therefore, there are many directions to proceed by generalizing the analysis in this paper. It would be interesting to see how the viscosities affect the instability. The nonlinear dynamics would be interesting too, even within the perfect-fluid approximation.

\subsection*{Acknowledgments}

The author would like to thank J.\ Camps, R.\ Emparan, N.\ Haddad, T.\ Harada, and N.\ Shibazaki for stimulating discussions, and anonymous referees for quite useful suggestions. This work was supported by Research Center for Measurement in Advanced Science in Rikkyo University, and by the Grant-in-Aid for Scientific Research Fund of the Ministry of Education, Culture, Sports, Science and Technology, Japan [Young Scientists (B) 22740176].

\appendix

\section{Comments on $r_{min}$}
\label{sec:r_min}

After submitting a draft of this paper, anonymous referees gave the author several useful comments and suggestions about the values of $r_{min}$ in actual physical systems. Here, some of these suggestions are noted for further studies.

\paragraph{General order estimate.} On dimensional basis, one can argue that the critical radius $r_{min}$ is of the order of a microscopic length scale as follow. Let us consider the relativistic case for simplicity. In many physical systems, it would be possible to assume $c_s^2 = dp_0/d\epsilon_0 \sim 1$, which is equivalent to assume that the compressibility is proportional to the inverse of the energy density, $\epsilon_0^{-1}d \epsilon_0/dp_0 \sim \epsilon_0^{-1}$. In this case, $r_{min}$ in equation \eqref{rel_crit2} is reduced to $r_{min} \sim \sigma/\epsilon_0$. Here, let us assume further that the surface tension is microscopically proportional to the energy density in the bulk as $\sigma \sim l \epsilon_0$, where $l$ is a microscopic length scale. Such a length scale $l$ could be a mean inter-molecular distance\footnote{This should be justified in a molecular theory of capillary force. See, \eg, \cite{Rowlinson}.}. With these assumptions, one obtains $r_{min} \sim l$, which suggests that the instability is irrelevant since the hydrodynamic description itself breaks down at the scale of $l$.

\paragraph{Liquid-drop model of nucleus.} It is widely known that many features of nucleus such as the global behavior of binding energy, surface oscillations, and nuclear fission, can be understood with the {\it liquid-drop model}, in which the nucleus is modeled by a liquid drop of an incompressible (at leading order) fluid~\cite{Bohr1,Bohr2}. However, the compressibility, which allows radial oscillations of a drop (\ie, the so-called breathing mode), is important since it is directly related to the equation of state of nuclear matter, and necessary to the accurate estimate of nuclear properties (radii, masses, giant resonances, \etc). Though the argument in the above paragraph seems to show the instability found in this paper is irrelevant to nuclei, it would be interesting to compare systematically the parameters in the liquid-drop model and those in this paper. Incidentally, the confine/deconfine phase transition in the quantum chromodynamics (QCD) is expected to be of the first order, and so the deconfined phase, \ie, {\it the quark-gluon plasma} (QGP), could exist as a drop of fluid around the critical temperature. Thus, it would be also interesting to consider the effect of compressibility to the QGP balls.

\paragraph{Granular matter.} Recently, it was reported that a kind of {\it granular matters} such as glass beads of tiny radius exhibit effective compressibility~\cite{granular1} and surface tension~\cite{granular2}\footnote{The existence of surface tension (capillarity) in a granular matter might be surprising since the attractive force between grains is much smaller than other forces at play (gravity, friction, inelasticity). In fact, the surface tension of the granular matter in experiment~\cite{granular2} (grass beads) stems from not the attractive force between the beads but a strong interaction between the beads and surrounding air.}. The dynamics of granular matters cannot be described by hydrodynamics in general, and furthermore the origin of the surface tension is different from that in fluids. Therefore, one cannot apply the result in this paper as is to granular matters. It would be interesting, however, to examine the possibility that the new capillary instability found in this paper or its analog is effective or not in granular matters. As far as the author knows, the sound velocity of granular matter could be relatively small (\eg, $c_s \sim 1 \;{\rm m/s}$ in an experiment~\cite{granular3}). Thus, there remains the possibility that the $r_{min} \propto c_s^{-2}$ could be large.

\section{Stability of drops for non-spherical perturbations}
\label{sec:non-s-pert}

\subsection{Non-relativistic case}
\label{sec:non-s}

We show the stability of non-relativistic droplets in ${\mathbb R^{1,n+1}}$ ($n \geq 1$) for non-spherical perturbations [and the stability of cylinder in ${\mathbb R^{1,n+2}}$ for homogeneous (in the $r$-direction) but non-spherical perturbations].

We work in the polar coordinates in which the line element of the flat space is given by
\be
	g_{IJ} d x^I d x^J
	=
	d r^2 + r^2 \gamma_{ij}(\theta) d\theta^i d\theta^j.
\ee
In this coordinates, continuity equation \eqref{cont} and Euler equation \eqref{euler} in the $r$- and $\theta^i$-directions are
\begin{align}
	( \pd_t + v^r \pd_r + v^i \pd_i ) \rho
	+
	\rho \left( \pd_r v^r + \frac{n}{r}v^r + D_i v^i \right)
	&=
	0,
\label{cont_non-s}
\\
	\rho
	\Big(
		( \pd_t + v^r \pd_r + v^i \pd_i )v^r
		- r \gamma_{ij} v^i v^j
	\Big)
	&= 
	-\pd_r p,
\label{euler_non-s_r}
\\
	\rho
	\left(
		\pd_t + v^r \pd_r + \frac{v^r}{r} + v^j D_j
	\right) v^i
	&=
	- \frac{\gamma^{ij}}{r^2} \pd_j p,
\label{euler_non-s_i}
\end{align}
where $D_i$ is the covariant derivative compatible with $\gamma_{ij}$. If we parametrize the scalar function $f$ as
\be
	f(t,r,\theta) = r - R (t,\theta_1,\theta_2,\ldots,\theta_n),
\ee
the mean curvature, appearing in Young-Laplace relation~\eqref{YL}, reads
\be
	\kappa
	=
	\frac{n}{R [ 1+R^{-2}(DR)^2 ]^{1/2}}
	-
	\frac{ R^2 D^2 R + ( D^2 R - R ) (DR)^2 - (D^i R) (D^j R) D_i D_j R }
		 { R^4 [ 1+R^{-2} (DR)^2 ]^{3/2} }.
\label{kappa_non-s}
\ee
Here, $(DR)^2 := \gamma^{ij} (D_i R) D_j R$ and $D^2 R := \gamma^{ij} D_i D_j R$. Kinetic boundary condition \eqref{kbc} is
\be
	\pd_t R + v^i \pd_i R = v^r \; \big|_{r=R}.
\label{kbc_non-s}
\ee

Obviously, equations of motion \eqref{cont_non-s}, \eqref{euler_non-s_r}, and \eqref{euler_non-s_i}, and boundary conditions \eqref{YL} [with equation \eqref{kappa_non-s}] and \eqref{kbc_non-s} allow the spherical drop as a static equilibrium, where the constant pressure $p_0$ and radius of sphere $R=r_0$ satisfy equation~\eqref{equi}. Now, we consider linear perturbations of this equilibrium resulting from the disturbance of surface,
\be
	R (t,\theta)
	=
	r_0 [ 1+\varepsilon e^{\omega t} Y(\theta) ].
\label{delta_non-s_h}
\ee
Here, $Y(\theta_1,\theta_2,\ldots,\theta_n)$ is the harmonic function on the unit $n$-sphere,
\be
	\big[ D^2 + \ell ( \ell+n-1 ) \big] Y(\theta) = 0,
\;\;\;
	\ell = 0,1,2,\ldots \; .
\ee
The perturbed pressure to $O(\varepsilon)$ may be written as
\be
	p(t,r,\theta) = p_0 [ 1 + \varepsilon e^{\omega t} P (r)Y(\theta) ].
\label{delta_non-s_p}
\ee
Substituting expression~\eqref{delta_non-s_p} into wave equation~\eqref{wave_eq}, we obtain
\be
	\frac{d^2 P}{ dr^2}
	+
	\frac{n}{r}\frac{  dP}{dr}
	-
	\left(
		\frac{\omega^2}{c_s^2} + \frac{ \ell(\ell + n-1) }{ r^2 }
	\right) P = 0.
\label{F_eq_non-s}
\ee
With the regularity at the origin and the perturbed Young-Laplace formula $\delta p = \sigma \delta \kappa |_{r=R}$, one finds the following solves equation \eqref{F_eq_non-s},
\be
	P (r)
	=
	\frac{ (\ell-1)(\ell+n) r_0^{(n-1)/2} }{ n I_{\ell + (n-1)/2} ( \omega r_0/c_s ) }
	\frac{ I_{\ell + (n-1)/2} \left( \omega r /c_s \right) }{r^{(n-1)/2}}.
\label{F_non-s}
\ee
Substituting equations~\eqref{equi}, \eqref{delta_non-s_h}, \eqref{delta_non-s_p}, and \eqref{F_non-s} into \eqref{delta_ph}, we obtain
\be
	\omega^2
	=
	-
	\frac{\sigma}{\rho_0 r_0^3}
	(\ell-1)(\ell+n)
		\left(
			\ell
			+
			\frac{ \omega r_0 }{ c_s } 
			\frac{ I_{\ell + (n+1)/2}(\omega r_0/c_s) }{ I_{\ell + (n-1)/2}(\omega r_0/c_s) }
		\right).
\label{disp_non-s}
\ee
If one takes $n=2$, $c_s \to \infty$, and writes $\omega \to {\rm i}\Omega $ ($ {\rm i} = \sqrt{-1}$) in this relation, one reproduces the classic formula~\eqref{drop-osc}, \ie, the angular frequency of oscillations for the incompressible inviscid droplet immersed in the three-dimensional vacuum~\cite{RayleighSound}.

If equation~\eqref{disp_non-s} has a positive root for a given $\ell$ ($\geq 0$), the spherical droplet is unstable for the perturbation labeled by $\ell$. Introducing the dimensionless quantities as in equation~\eqref{dimless}, the problem is equivalent to find positive zero of the following function,
\be
	F_\ell (\hat{\omega})
	:=
	\hat{\omega}^2
	+
	(\ell-1)\ell(\ell+n)
	+
	(\ell-1)(\ell+n) \beta \hat{\omega}
	\frac{ I_{\ell + (n+1)/2} (\beta \hat{\omega}) }{ I_{\ell + (n-1)/2} ( \beta \hat{\omega} ) }.
\label{G_omega}
\ee
Note that $F_0(\hat{\omega})$ is nothing but $F(\hat{\omega},0)$ in section~\ref{sec:k=0}, which was shown to have the positive zero. Taking into account the positivity of the modified Bessel function, one can easily see
\be
	F_{\ell} (\hat{\omega}) > 0
\;\;\;
	\mbox{for}
\;\;\;
	\ell \geq 1
\;\;\;
	\mbox{and}
\;\;\;
	\hat{\omega} > 0.
\label{G_1}
\ee
Thus, there is no positive zero of $F_\ell(\hat{\omega})$ for $\ell \geq 1$, proving the stability of droplet for the non-spherical perturbations.

\subsection{Relativistic case}
\label{sec:non-s_rel}

We show the stability of relativistic droplets in ${\mathbb R^{1,n+1}}$ ($n\geq 1$) for the non-spherical perturbations [and that of relativistic cylinders in ${\mathbb R^{1,n+2}}$ for non-spherical but homogeneous ($k=0$) perturbations].

We write the flat metric as
\be
	g_{\mu\nu} d x^\mu d x^\nu
	=
	- d t^2 + d r^2 + r^2 \gamma_{ij} d \theta^i d \theta^j.
\ee
Then, we perturb the spherically symmetric static equilibrium, where the constant pressure $p_0$ and radius $R=r_0$ satisfy equation~\eqref{equi}, with the same ansatz \eqref{delta_non-s_h} and \eqref{delta_non-s_p} as the non-relativistic case. The mean curvature \eqref{kappa_rel} reads
\begin{multline}
	\kappa
	=
	\frac{ n }{ R [ 1-(\pd_t R)^2 + R^{-2} (DR)^2 ]^{1/2} }
	+
	\frac{1}{ R^4 [ 1-(\pd_t R)^2 + R^{-2} (DR)^2 ]^{3/2} }
	\Big(
		[ R^2 + (DR)^2 ] R^2 \pd_t^2 R
\\
		- [ 1-(\pd_t R)^2 ] R^2 D^2 R
		+ ( R-D^2 R )( DR )^2
		+ (D^i R) (D^j R) D_i D_j R
	\Big).
\end{multline}

Substituting expression \eqref{delta_non-s_p} into wave equation \eqref{wave_eq_rel}, one finds that radial function $P(r)$ satisfies the same form of equation as \eqref{F_eq_non-s}. With the perturbed kinetic boundary condition, one can fix an integration constant to obtain
\be
	P(r)
	=
	\frac{ [ (\omega r_0)^2 + (\ell-1)(\ell+n) ]r_0^{(n-1)/2} }{ n I_{\ell + (n-1)/2} (\omega r_0/c_s) } 
	\frac{ I_{\ell + (n-1)/2} (\omega r/c_s) }{ r^{(n-1)/2} }.
\label{F_non-s_rel}
\ee
Substituting equations~\eqref{delta_non-s_h}, ~\eqref{delta_non-s_p}, and \eqref{F_non-s_rel} into \eqref{delta_hp_rel}, one obtains
\be
	\omega^2
	=
	- \frac{ \sigma }{ ( \epsilon_0 + p_0 )r_0^3 }
	[ (\omega r_0)^2 + (\ell-1)(\ell+n) ]
	\left(
		\ell + \frac{\omega r_0}{c_s} \frac{ I_{\ell + (n+1)/2}(\omega r_0/c_s) }{ I_{\ell + (n-1)/2}(\omega r_0/c_s) }
	\right).
\ee

In terms of the dimensionless quantities in equation~\eqref{dimless_rel}, to find a positive root of the above equation is equivalent to find a positive zero of the following function, 
\be
	F_\ell (\hat{\omega})
	:=
	\hat{\omega}^2
	+
	\frac{ \hat{\sigma} }{ 1+n\hat{\sigma} }
	[ \hat{\omega}^2 + (\ell-1)(\ell+n) ]
	\left(
		\ell + \beta \hat{\omega} \frac{ I_{\ell + (n+1)/2}(\beta\hat{\omega}) }{ I_{\ell + (n-1)/2}(\beta\hat{\omega}) }
	\right).
\label{G_ell_rel}
\ee
Observe that $F_0(\hat{\omega})$ is nothing but $F(\hat{\omega},0)$ in section~\ref{sec:k=0-rel}, which was shown to have the positive zero. One can easily show with using the positivity of the modified Bessel functions that
\be
	F_{\ell} (\hat{\omega}) > 0
\;\;\;
	\mbox{for}
\;\;\;
	\ell \geq 1,
\;\;\;
	\hat{\sigma} >  0,
\;\;\;
	\mbox{and}
\;\;\;
	\hat{\omega} > 0.
\ee
Thus, $F_\ell(\hat{\omega})$ for $\ell \geq 1$ has no positive root, proving the stability of relativistic droplets for the non-spherical perturbations.



\end{document}